\documentclass[twocolumn,twoside,preprintnumbers,amsmath,amssymb,pacs]{revtex4}
\usepackage{epsfig}
\usepackage{graphicx}
\usepackage{dcolumn}
\usepackage{bm}

\usepackage{fancyhdr}

\usepackage{pslatex}

\newcommand{\be}{\begin{eqnarray}}
\newcommand{\ee}{\end{eqnarray}}

\begin{document}

\title{{\Large Connecting an effective model of confinement and chiral 
symmetry to lattice QCD}}

\author{E. S. {\sc Fraga}$^1$\footnote{fraga@if.ufrj.br} 
and \'A. {\sc M\'ocsy}$^2$$^3$\footnote{mocsy@bnl.gov}}
\affiliation{
$^1$Instituto de F\'\i sica, 
Universidade Federal do Rio de Janeiro, 
C.P. 68528, Rio de Janeiro, RJ 21941-972, Brazil \\
$^2$RIKEN-BNL Research Center, Brookhaven National Laboratory, 
Upton, New York 11973, USA\\
$^3$Frankfurt Institute for Advanced Studies, Max von Laue Str. 1, 60438 Frankfurt am Main, Germany
}


\begin{abstract}

We construct an effective model for the chiral field and the Polyakov loop 
in which we can investigate the interplay between the approximate chiral 
symmetry restoration and the deconfinement of color in a thermal SU(3) gauge 
theory with three flavors of massive quarks. 
The phenomenological couplings between these two sectors can then be related 
to the recent lattice data on the renormalized Polyakov loop and the chiral
condensate close to the critical region. 
\vspace{0.5cm}

PACS numbers:

Keyword: Confinement; Chiral phase transition; Quark mass effects
\end{abstract}


\maketitle

\thispagestyle{fancy}
\setcounter{page}{0}

\section{Introduction}

Various theoretical developments accompanied by several numerical results from 
lattice simulations have advanced our understanding of QCD thermodynamics \cite{{Petreczky:2006zk},Bernard:2006xg}. Nevertheless, there are still some important 
qualitative questions that remain to be answered. For instance, as one increases the 
temperature above the critical one, which mechanism drives the QCD phase transition for a 
given quark mass? Is the QCD transition more deconfining or chiral symmetry restoring? Although 
lattice QCD might provide definitive responses soon, some current findings are still 
controversial.  In particular, the 
numerical value of the critical temperature for the QCD phase transition \cite{Petreczky:2004xs},  
the uncertainty on the location and even the possible absence of a critical point in the 
temperature-baryon density plane \cite{deForcrand:2006pv}, and the newly  suggested 
possible difference between the critical temperatures for the chiral and and the deconfinement 
transitions \cite{Aoki:2006br}. Therefore, it is useful to address these issues from a 
phenomenological point of view, resorting to effective field theory models.

The pure gauge sector of QCD with $N$ colors, corresponding to infinitely heavy quarks, 
is well under control in lattice simulations, with a precise prediction for the deconfinement 
critical temperature, and a good understanding of its thermodynamic 
behavior \cite{Laermann:2003cv}. This sector has a global $Z_N$ symmetry associated 
with the center of the gauge group $SU(N)$. The Polyakov loop $\ell$,
charged under $Z_N$, serves as order parameter for the deconfining
transition. It is real for $N=2$, and the $Z_2$ symmetry
breaking deconfinement transition is of second order \cite{Damgaard:1987wh}. For $N=3$, 
$\ell$ is complex and transforms as $\ell\rightarrow e^{2\pi i/3} \ell$ \cite{Pisarski:2002ji}. 
Accordingly, the expectation value $\langle\ell\rangle$ vanishes at low
temperatures, when $Z_3$ is unbroken, and $\langle\ell\rangle\neq 0$ 
above the deconfinement critical temperature $T_d\sim 270~$MeV. 
This transition is verified to be weakly first order \cite{Fukugita:1983ge}.  

Another theoretically well-studied sector of QCD is the limit of zero and
small quark masses, i.e. the chiral limit. In this regime $Z_N$ is always 
broken, but chiral symmetry is restored above a critical temperature $T_c$.
The order parameter is the quark chiral condensate $\sigma$. The
transition is believed to be of second order for two \cite{Pisarski:1983ms,note}, 
and of first order for three flavors of massless quarks. For finite quark masses 
chiral symmetry is explicitly broken. Increasing $m_q$ from zero corresponds to 
weakening the first-order phase transition. For some quark mass the
first-order critical line turns into a second-order chiral
critical point. For even larger $m_q$ the transition becomes a 
crossover. Effective field theories have been used to analyze 
this sector \cite{Scavenius:2000qd}.

Deconfinement and chiral symmetry restoration are phenomena
of very different nature, corresponding to approximate symmetries 
of the QCD Lagrangian. For realistic quark masses both 
chiral and $Z_N$ symmetries are explicitly broken, and most likely 
both transitions are in the crossover domain \cite{Gavin:1993yk}. The phase diagram 
illustrating the quark mass dependence of the deconfinining and the chiral symmetry 
restoring transition has been qualitatively 
investigated in effective theories in \cite{Gavin:1993yk,{Dumitru:2003cf},{Mocsy:2004nn}} 
and on the lattice in \cite{Karsch:2000kv,{Karsch:2003va},{Philipsen:2005mj}}.
In the crossover region neither $\sigma$ nor $\ell$ can play the role of a true
order parameter. However, lattice simulations show that the
susceptibilities associated with these two quantities peak at the
same temperature when quarks are in the fundamental
representation of the gauge group \cite{Karsch:1998qj}, suggesting
that the transitions occur at the same critical temperature.
Similar results were obtained also in terms of the chemical
potential \cite{Alles:2002st}. Recent, more refined results confirm the coincidence of 
the two transitions \cite{Cheng:2006qk}, although results using a different 
method find critical temperatures that differ by 
$\sim 25~$MeV \cite{Aoki:2006br}, showing that this issue is still not settled.

There exists quite a substantial effort to understand the true nature of these transitions 
using different effective field 
theories 
\cite{Meisinger:2002kg,{Digal:2000ar},{Hatta:2003ga},{Fukushima:2003fw},{Mocsy:2003qw},{Megias:2004hj},{Ratti:2005jh}}. 
We especially consider Ref. \cite{Mocsy:2003qw}, in which the authors provided an explanation, 
within a generalized Ginzburg-Landau theory, to why for $m_q=0$ chiral
symmetry restoration leads to deconfinement. The analysis was
based on the following general idea: the behavior of an order
parameter induces a change in the behavior of non-order parameters
at the transition \cite{Mocsy:2003tr} via the presence of a
possible coupling between the fields, $g_1\ell\sigma^2$. In
\cite{Mocsy:2003qw} it was assumed that $g_1>0$, which is in line 
with recent results from the lattice.

In this paper we extend the analysis of \cite{Mocsy:2003qw} to the case 
of nonzero quark masses. Within this model we can then not only study the 
qualitative relationship between deconfinement and chiral symmetry restoration 
including massive quarks, but also discuss how to provide a few semi-quantitative 
predictions for the behavior of the condensates and critical temperatures adopting 
values for the couplings constrained by lattice data. 

Lattice QCD results for the temperature dependence of the chiral condensate for three 
degenerate flavors of massive quarks were reported in \cite{Bernard:2003dk}, and the 
first results for the behavior of the renormalized Polyakov loop under the same lattice 
conditions were presented in \cite{Petreczky:2004pz}. Most recent lattice data, using 
improved lattice actions, provide the temperature dependence of the chiral 
condensates \cite{Schmidt:2006ku,{Schmidt:2006vz}} and the Polyakov loop for 2+1 
flavors \cite{{Schmidt:2006vz},Petrov:2006pf}. They indicate that the chiral and 
deconfinement transitions coincide but, contrary to the case of the chiral condensate, the 
flavor and quark mass dependence of the Polyakov loop is contained entirely in the  flavor 
and quark mass dependence of the critical temperature. In \cite{Schmidt:2006ku} the 
strange quark condensate was studied separately from the light quark condensate, and 
it was found that chiral symmetry restoration happens together for the quarks of different 
flavors. Furthermore, it was shown in \cite{Schmidt:2006vz} that, as expected, the strength 
of the chiral transition decreases while that of the deconfining transition increases with 
increasing quark mass. 

This paper is organized as follows. In Section II we describe the 
effective theory in which we couple the Polyakov loop to the chiral sector. 
In Section III we discuss how to relate our model to the recent findings from the 
lattice in order to constraint the couplings of our model. Here we set up the theory. 
We will report on the quantitative analysis elsewhere \cite{next}. 
Section IV contains our summary and outlook. 

\section{The Effective Theory}

For nonzero quark masses both the chiral and the $Z_N$ symmetries are
explicitly broken, so that neither the chiral condensate nor the Polyakov
loop play the role of a true order parameter any longer. One can, however, still use these 
quantities to monitor the approximate transitions, as is done in lattice QCD calculations, 
where the transition temperatures are determined from the peak position of the corresponding 
susceptibilities. Using these quantities we can thus construct an effective field theory to study 
the interplay between the approximate chiral symmetry restoration and the deconfinement of 
color. In what follows, we write down the most general renormalizable effective Lagrangian 
that can be built from the chiral field $\sigma$ and the Polyakov loop $\ell$, following the form 
introduced in \cite{Mocsy:2003qw}.   

For the chiral part of the potential we chose the $SU(3)\times SU(3)$ linear sigma model
\be
V_\chi(\Phi)&=&m^2\mbox{Tr}[\Phi^{\dagger}\Phi] +
\lambda_1(\mbox{Tr}[\Phi^{\dagger}\Phi])^2 +
\lambda_2\mbox{Tr}[(\Phi^{\dagger}\Phi)^2] \nonumber \\
&-&c(\det[\Phi]+\det[\Phi^\dagger]) -
\mbox{Tr}[H(\Phi^\dagger+\Phi)]\, ,
\label{su3}
\ee
where $\Phi=T_a\phi_a=T_a(\sigma_a+i\pi_a)$ is a complex $3\times 3$
matrix containing the scalar $\sigma_a$ and the pseudoscalar
$\pi_a$ fields, and $H=T_ah_a$ is the explicit symmetry breaking
term. Here $T_a=\lambda_a/2$ are the nine generators of $SU(3)$
with $\lambda_a$ the Gell-Mann matrices, 
$\lambda_0=\sqrt{2/3}\bf{1}$, and $h_a$ nine external
fields. The different symmetry breaking patterns of this theory
have been thoroughly investigated in \cite{Lenaghan:2000ey} and
\cite{Roder:2003uz}. We are interested in making a direct
comparison with lattice results both for the $N_f=3$ 
degenerate mass flavors, $m_u=m_d=m_s$, and for the more realistic 
$N_f=2+1$ case, with $m_u=m_d<m_s$. 

Here we set up the degenerate $N_f=3$ case, for which the potential (\ref{su3}) 
undergoes some simplifications. This particular case corresponds to that named 
4a in \cite{Lenaghan:2000ey}. Accordingly, $c=0$, since
the $U(1)_A$ anomaly is not taken into account, and $h_0\neq 0$ is
the only symmetry breaking term, since $h_3=h_8=0$. In this case, 
the scalar and pseudoscalar matrices are diagonal and the masses
are given by 
\be
m_\sigma^2&=&m^2+3(\lambda_1+\frac{\lambda_2}{3})\sigma_0^2\, ,
\nonumber\\
m_{f_0}^2&=&m^2+(\lambda_1+\lambda_2)\sigma_0^2\, ,\\
m_\pi^2 & =&m^2+(\lambda_1+\frac{\lambda_2}{3})\sigma_0^2\, .
\nonumber
\ee 
In the vacuum, the condensate is $\sigma_0=\sqrt{2/3}f_\pi$ and
$h_0=\sqrt{2/3}f_\pi m_\pi^2$, thus $\Phi=f_\pi$ and
$H=f_\pi m_\pi^2$.

The idea now is to write this special case of the $SU(3)$ potential in the form 
\be
V_{\chi}(\sigma) = \frac{m^2}{2}\sigma^2 +
\frac{\lambda}{4}\sigma^4 - H\sigma\, . 
\label{sigmapot}
\ee 
This choice is motivated by the fact that the up-down condensate is
aligned in the $\sigma$ direction. Furthermore, without the
$U(1)_A$ anomaly the difference between the results obtained for 
$SU(2)$ and $SU(3)$ in the sigma model are negligibly small. This is
clearly illustrated in the melting of the condensates shown on Fig. 6 
from \cite{Roder:2003uz}. In (\ref{sigmapot}) the coupling is given by 
$\lambda=\lambda_1+\lambda_2/3$. The parameters of the
model $m^2$, $\lambda_1$ and $\lambda_2$ are determined from the vacuum
values of $m_\sigma$, $m_\pi$ and $m_{f_0}$. In what follows we assume that 
pions and $f_0$'s will not play a role, so we can discard them from the
outset. 

We adopt a mean field analysis, and thus replace the fields by their expectation values. 
Then, as customary, we
choose the vacuum to be aligned in the sigma direction. As a result, 
all our relevant equations will depend only on
$\langle\sigma\rangle$. In the $\sigma$ direction the potential
has the form shown in Eq. (\ref{sigmapot}).

The Polyakov loop potential when the $Z_3$ symmetry is explicitly broken reads 
\be
V_{PL}(\ell)=g_0\frac{\ell+\ell^*}{2}+\frac{m_0^2}{2}|\ell|^2
-\frac{g_3}{3}\frac{\ell^3+\ell^{*3}}{2} +
\frac{g_4}{4}(|\ell|^2)^2\, .
\label{ellpot}
\ee
Effective field theories using this potential have been introduced 
in \cite{Pisarski:2000eq} and further analyzed in \cite{Dumitru:2000in}. 
Here we focus only on the real part of the Polyakov loop, 
since one can always choose the expectation value to be real, at 
least for $\mu=0$. In this case, the contributions coming from the 
Polyakov loop potential simplify to
\be
V_{PL}(\ell) = g_0\ell + \frac{m_0^2}{2}\ell^2 +
\frac{g_3}{3}\ell^3 + \frac{g_4}{4}\ell^4 \, .
\ee
Here, $\ell$ is a scalar field (with dimensions of mass)
proportional to the Polyakov loop (a dimensionless quantity), and
$m_0$ is its bare mass above the transition.  The couplings $g_i$
that appear in each piece of the complete potential can be fixed
by lattice data as will be shown in the next section. 

The interactions between the chiral field and the Polyakov loop which exist 
when both symmetries are spoiled by the presence of massive quarks yield 
the following contribution to the total effective potential 
\be
V_{int}(\ell,\Phi) &=&
(g_1\ell+g_2\ell^2)\mbox{Tr}[\Phi^{\dagger}\Phi] \nonumber  \\
&+&(\bar{g}_1\ell^2 
+ \bar{g}_2\ell)\mbox{Tr}[H(\Phi^\dagger+\Phi)] \, ,
\ee
which simplifies to
\be
V_{int}(\ell,\sigma) = g_1\ell\sigma^2 + g_2\ell^2\sigma^2 +
\bar{g}_1\ell^2\sigma + \bar{g}_2\ell\sigma\, .
\ee
We now expand the fields around their mean values,
$\sigma=\langle\sigma\rangle+\delta\sigma$ and
$\ell=\langle\ell\rangle+\delta\ell$, and compute the mean-field
equations of motion from first variations of the complete effective 
potential $V=V_{PL}+V_{\chi}+V_{int}$, obtaining:
\be
2\lambda\langle\sigma\rangle^3 - m_\sigma^2\langle\sigma\rangle +
H - \bar{g}_1\langle\ell\rangle^2 - \bar{g}_2\langle\ell\rangle = 0
\ee
and
\be
g_0 + m_\ell^2\langle\ell\rangle - g_3\langle\ell\rangle^2 -
2g_4\langle\ell\rangle^3 + g_1\langle\sigma\rangle^2 +
\bar{g}_2\langle\sigma\rangle = 0 \, .
\ee
The mass of the $\sigma$ field, $m_\sigma$, is determined
as the second derivative of the potential with respect to
$\sigma$ evaluated at the minimum:
\be
m_\sigma^2 = m^2 + 3\lambda\langle\sigma\rangle^2 +
2g_1\langle\ell\rangle + 2g_2\langle\ell\rangle^2 \, .
 \ee
Similarly, the mass of the Polyakov loop is given by
\be
m_\ell^2 = m_0^2 + 2g_3\langle\ell\rangle +
3g_4\langle\ell\rangle^2 +2\bar{g_1}\langle\sigma\rangle +
2g_2\langle\sigma\rangle^2 \, .
\ee

>From now on, we neglect higher-order contributions from
$\langle\ell\rangle$, but not from $\langle\sigma\rangle$.
We have to keep higher-order terms in the chiral field to be
able to recover the correct chiral limit discussed in
Ref. \cite{Mocsy:2003qw},
when quarks are massless and $H=0$. Moreover, couplings which
were included in the interaction potential of the present analysis due
to explicit chiral symmetry breaking must vanish for $H=0$. For this
reason, we require these couplings to be linear in
$H \sim m_q + {\cal{O}}(m_q^2)$. Accordingly, we make the following
replacements: $\bar{g}_1\rightarrow \bar{g}_1H$ and
$\bar{g}_2\rightarrow \bar{g}_2H$. The equations of motion thus
become
\be
2\lambda\langle\sigma\rangle^3-m_\sigma^2\langle\sigma\rangle +
H - \bar{g}_2H\langle\ell\rangle = 0
\ee
and
\be
g_0 + m_\ell^2\langle\ell\rangle + g_1\langle\sigma\rangle^2 +
\bar{g}_2H\langle\sigma\rangle = 0 \, .
\ee

To first order in $m_q$, we have:
\be
\langle\ell\rangle &\simeq& - \frac{g_0}{m_\ell^2} -
\frac{g_1}{m_\ell^2}\langle\sigma\rangle^2 -
\frac{\bar{g}_2}{m_\ell^2}H\langle\sigma\rangle
\label{expectation-l}
\ee
and
\be
\langle\sigma\rangle^3 &+& \frac{g_1\bar{g}_2}{2\lambda
m_\ell^2}H\langle\sigma\rangle^2 -
\frac{m_\sigma^2}{2\lambda}\langle\sigma\rangle \nonumber \\
&+& \frac{H}{2\lambda}\left(1+ \frac{g_0\bar{g}_2}{m_\ell^2}\right)
\simeq 0 \, .
\label{eom-sigma}
\ee
Notice that in the chiral limit, i.e. for $H=0$, we recover the
results from Ref. \cite{Mocsy:2003qw}. One can now use
(\ref{expectation-l}) and (\ref{eom-sigma}) to extract the
couplings from lattice results for the renormalized Polyakov loop
and the chiral condensate in the presence of massive quarks. In 
the following we discuss how that can be implemented in a 
general fashion \cite{next}.

\section{Connecting to lattice data}

Our aim is to use the above derived equations to describe the corresponding 
lattice data for the chiral condensate \cite{Bernard:2003dk} and the Polyakov 
loop \cite{Petreczky:2004pz}  for three degenerate flavors, generated under the 
same lattice conditions. We can apply the same procedure for the 2+1 flavor 
most recent lattice data \cite{{Schmidt:2006vz},Schmidt:2006ku,{Petrov:2006pf}}. 
In this case we also study separately the light and strange quark 
condensates \cite{next}. 

In order to compare our mean-field results to lattice data, we use
the Gell-Mann--Oakes--Renner (GOR) relation
\be
m_q\langle\bar{\psi}\psi\rangle = -f_\pi
m_\pi^2\langle\sigma\rangle \, ,
\ee
which in the vacuum yields $m_q\langle\bar{\psi}\psi\rangle_0 =
-f_\pi^2 m_\pi^2$, and define the dimensionless quantity that is
measured on the lattice
$x\equiv\langle\bar{\psi}\psi\rangle/\langle\bar{\psi}\psi\rangle_0$,
where we take $\langle\bar{\psi}\psi\rangle_0=-2(225~$MeV$)^3$.
The mass of the sigma meson is given by
\be
m_\sigma^2 (T,m_q) = m_\pi^2(m_q) + 2\lambda
\langle\sigma\rangle^2(T) \, .
\ee
 Furthermore, as customary in a Ginzburg-Landau theory, we assume
\be
m_\sigma^2 (T,m_q=0) = a T_c^2\left(1-\frac{T}{T_c}\right)\, , 
\ee
below the transition, where $a$ is a dimensionless constant. This
is characteristic of a second-order phase transition. For nonzero
quark mass the transition is a crossover. We then 
consider that the temperature and quark mass dependence of the
sigma mass are well described by the following ansatz
\be
m_\sigma^2 (T,m_q) = m_\pi^2(m_q) + a
T_c^2\left(1-\frac{T}{T_c}\right)\, . 
\ee

To describe the temperature-dependence of the Polyakov loop mass, 
a fit to pure SU(3) Yang-Mills results from the lattice
\cite{Kaczmarek:1999mm} was performed in \cite{Dumitru:2000in}.
Accordingly: 
\be 
m_\ell^2 (T) =\left(\frac{s(T)}{T}\right)^2\, ,
\ee 
where the temperature-dependent string tension is given by
$s(T)=1.21\sqrt{s_0^2-0.99T^2\sigma_0/(0.41)^2}$, with the
zero-temperature string tension $s_0=(440~$MeV$)^2$. This parametrization, 
however, is not valid in full QCD. Therefore we keep the Polyakov loop mass 
as a free parameter. 

Given the conditions above, we can rewrite
(\ref{eom-sigma}) in a more convenient form:
\be
x^3 &+& C_2 \left[ \frac{g_1\bar{g}_2}{m_\ell^2(T)} \right]  m_q x^2
- C_1 m_\sigma^2(T,m_q) ~x \nonumber \\
&+& C_0 \left[ 1 +
\frac{g_0\bar{g}_2}{m_\ell^2(T)} \right] m_q = 0 \, ,
\label{eom-x} 
\ee
where we have defined the following constants: 
$C_2\equiv -\langle\bar{\psi}\psi\rangle_0/2\lambda f_\pi^2
\simeq 33~$MeV,
$C_1\equiv 1/2\lambda f_\pi^2
\simeq 2.9 \times 10^{-6}~$MeV$^{-2}$ and
$C_0\equiv -\langle\bar{\psi}\psi\rangle_0/2\lambda f_\pi^4
\simeq 3.8 \times 10^{-3}~$MeV$^{-1}$.

We can also define the dimensionless quantity
\be
y &\equiv& \frac{\langle\ell\rangle}{(-g_0/m_\ell^2(T))}
\nonumber \\
&=& 1 - \frac{g_1}{(-g_0)}\langle\sigma\rangle^2 -
\frac{\bar{g}_2}{(-g_0)}H \langle\sigma\rangle \, ,
\ee
to be compared with lattice results for the renormalized
Polyakov loop as a function of temperature for different quark
masses \cite{Petreczky:2004pz}. Notice that $y$ has the
expected behavior in the limit of very high temperatures. In
terms of the dimensionless chiral ratio $x$ and the quark mass
$m_q$:
\be
y=1- \frac{g_1}{(-g_0)}f_\pi^2 x^2 +
\frac{\bar{g}_2}{(-g_0)}\langle\bar{\psi}\psi\rangle_0 m_q x \, .
\label{eom-y}
\ee

The functions $x=x(T)$ and $y=y(T)$ for a few values of $m_q$,
and also the corresponding values of $T_c$, were computed on the
lattice \cite{Bernard:2003dk,Petreczky:2004pz}. By fitting equations
(\ref{eom-x}) and (\ref{eom-y}), one can extract from the data the couplings
$a$, $g_0$, $g_1$ and $\bar{g}_2$, to be used in the effective 
field theory. Our preliminary results indicate that the sign of the coupling $g_1$ 
complies with the expectation of \cite{Mocsy:2003qw}, assuring that the chiral and 
the deconfinement transitions coincide. Detailed results with further conclusions 
are deferred to our upcoming publication \cite{next}. 

\section{Summary and outlook}

In this paper we have sketched a systematic method to construct a phenomenological 
generalized Ginzburg-Landau effective theory describing simultaneously the 
processes of chiral symmetry restoration and deconfinement in the presence of 
massive quarks. We devoted special attention to the behavior of the quasi order 
parameters $\sigma$ and $\ell$ with temperature, which can be connected 
to lattice data. The latter, on the other hand, can be used to provide constraints on 
the couplings of the effective theory. A detailed analysis of the method, as well as 
the extraction of the range of physical values of the couplings will be presented 
elsewhere \cite{next}.

The effective theory presented in this paper is very simple. 
For instance, if one studies the renormalization of Polyakov loops, one is naturally 
lead to consider effective {\it matrix} models for the deconfinement transition, 
which unfolds a rich set of possibilities, especially at large $N$ \cite{matrix}. In 
particular, eigenvalue repulsion from the Vandermonde determinant in the 
measure seems to play a key role \cite{Pisarski:2006hz}. These studies pointed out 
however, that in the neighborhood of the transition the relevant quantity is the trace 
of the Polyakov loop, underlying the relevance of our effective theory in this region. 
The simplicity of our model allows for a direct comparison to lattice data in a way 
that provides definite constraints on the couplings, after which one can test and 
predict semi-quantitatively the behavior of the approximate order parameters 
in different settings.

The full determination of the couplings of an effective model linking confinement 
and chiral symmetry breaking using reliable lattice data is useful for the 
study of the phase diagram for strong interactions, and can bring understanding 
to some of the open questions considered in the introduction. Furthermore, the effective 
potential derived from this model can be used in the study of the real-time dynamics 
of the QCD phase transitions, including the effects from dissipation and noise that 
result from self and mutual interaction of the fields related to the chiral condensate 
and the Polyakov loop \cite{dissipation}. Results from a real-time study within 
our effective model will be reported in the future \cite{combined}.

\acknowledgments
We thank A. Dumitru, P. Petreczky and R. Pisarski for  productive discussions, 
and the Service de 
Physique Th\'eorique (CEA-Saclay), where part of this work 
has been done, for their kind hospitality. When this work was initiated \'A.M. received 
financial support from the Alexander von Humboldt Foundation.
The work of E.S.F is partially supported by CAPES, CNPq, FAPERJ and FUJB/UFRJ.



\begin{thebibliography}{99}

\bibitem{Petreczky:2006zk}
  P.~Petreczky,
  hep-lat/0609040.
  
\bibitem{Bernard:2006xg}
  C.~Bernard {\it et al.},
  hep-lat/0610017.
  
\bibitem{Petreczky:2004xs}
  P.~Petreczky,
  Nucl.\ Phys.\ Proc.\ Suppl.\  {\bf 140}, 78 (2005).

\bibitem{deForcrand:2006pv}
  P.~de Forcrand and O.~Philipsen,
  hep-lat/0607017.

\bibitem{Aoki:2006br}
  Y.~Aoki, Z.~Fodor, S.~D.~Katz and K.~K.~Szabo,
  hep-lat/0609068.

\bibitem{Laermann:2003cv}
E.~Laermann and O.~Philipsen,
Ann.\ Rev.\ Nucl.\ Part.\ Sci.\  53 (2003) 163.

\bibitem{Damgaard:1987wh}
  P.~H.~Damgaard,
  Phys.\ Lett.\ B {\bf 194}, 107 (1987); 
  J.~Engels et al, 
  Phys.\ Lett.\ B {\bf 365}, 219 (1996).

\bibitem{Pisarski:2002ji}
  R.~D.~Pisarski,
  hep-ph/0203271.

\bibitem{Fukugita:1983ge}
  M.~Fukugita, T.~Kaneko and A.~Ukawa,
  Phys.\ Rev.\ D {\bf 28}, 2696 (1983);
  P.~Bacilieri {\it et al.},
  Phys.\ Rev.\ Lett.\  {\bf 61}, 1545 (1988); 
  F.~R.~Brown et al, 
  Phys.\ Rev.\ Lett.\  {\bf 61} (1988) 2058.
    
\bibitem{Pisarski:1983ms}
  R.~D.~Pisarski and F.~Wilczek,
  Phys.\ Rev.\ D {\bf 29}, 338 (1984).
  
\bibitem{note}
Lattice simulations, however, have not been able to settle
this issue, listed as an open problem in recent review talks in
the field (see, e.g., \cite{Heller:2006ub}). In particular, the transition might 
also be of first order. Perhaps the use of the present effective 
model in the two-flavor case could help to clarify this issue.

\bibitem{Heller:2006ub}
  U.~M.~Heller,
  hep-lat/0610114.

\bibitem{Scavenius:2000qd}
For instance: O.~Scavenius, A.~Mocsy, I.~N.~Mishustin and
D.~H.~Rischke,
Phys.\ Rev.\ C {\bf 64}, 045202 (2001);
  A.~Jakovac, A.~Patkos, Z.~Szep and P.~Szepfalusy,
  Phys.\ Lett.\ B {\bf 582}, 179 (2004).
  
\bibitem{Gavin:1993yk}
  S.~Gavin, A.~Gocksch and R.~D.~Pisarski,
  Phys.\ Rev.\ D {\bf 49}, 3079 (1994).

\bibitem{Dumitru:2003cf}
  A.~Dumitru, D.~Roder and J.~Ruppert,
  Phys.\ Rev.\ D {\bf 70}, 074001 (2004).

\bibitem{Mocsy:2004nn}
  \'A.~M\'ocsy,
  J.\ Phys.\ G {\bf 31}, S1203 (2005).

\bibitem{Karsch:2000kv}
  F.~Karsch, E.~Laermann and A.~Peikert,
  Nucl.\ Phys.\ B {\bf 605}, 579 (2001).
  
\bibitem{Karsch:2003va}
  F.~Karsch, C.~R.~Allton, S.~Ejiri, S.~J.~Hands, O.~Kaczmarek, E.~Laermann and C.~Schmidt,
  Nucl.\ Phys.\ Proc.\ Suppl.\  {\bf 129}, 614 (2004).
  
\bibitem{Philipsen:2005mj}
  O.~Philipsen,
  PoS {\bf LAT2005}, 016 (2006)
  [PoS {\bf JHW2005}, 012 (2006)].

\bibitem{Karsch:1998qj}
F.~Karsch and M.~Lutgemeier,
Nucl.\ Phys.\ B {\bf 550}, 449 (1999).

\bibitem{Alles:2002st}
B.~Alles, M.~D'Elia, M.~P.~Lombardo and M.~Pepe,
hep-lat/0210039.

\bibitem{Cheng:2006qk}
  M.~Cheng {\it et al.},
  Phys.\ Rev.\ D {\bf 74}, 054507 (2006).
  
\bibitem{Meisinger:2002kg}
  T.~R.~Miller and M.~C.~Ogilvie,
  Phys.\ Lett.\ B {\bf 488}, 313 (2000); 
  P.~N.~Meisinger, T.~R.~Miller and M.~C.~Ogilvie,
  Phys.\ Rev.\ D {\bf 65}, 034009 (2002); 
  P.~N.~Meisinger, T.~R.~Miller and M.~C.~Ogilvie,
  Nucl.\ Phys.\ Proc.\ Suppl.\  {\bf 119}, 511 (2003).
  
\bibitem{Digal:2000ar}
  S.~Digal, E.~Laermann and H.~Satz,
  Eur.\ Phys.\ J.\ C {\bf 18}, 583 (2001).
  
\bibitem{Hatta:2003ga}
  Y.~Hatta and K.~Fukushima,
  Phys.\ Rev.\ D {\bf 69}, 097502 (2004). 
  
\bibitem{Fukushima:2003fw}
  K.~Fukushima,
  Phys.\ Lett.\ B {\bf 591}, 277 (2004)

\bibitem{Mocsy:2003qw}
\'A.~M\'ocsy, F.~Sannino and K.~Tuominen,
Phys.\ Rev.\ Lett.\  {\bf 92}, 182302 (2004). 

\bibitem{Megias:2004hj}
  E.~Megias, E.~Ruiz Arriola and L.~L.~Salcedo,
  Phys.\ Rev.\ D {\bf 74}, 065005 (2006); 
  hep-ph/0610095;
  hep-ph/0610163.

\bibitem{Ratti:2005jh}
  C.~Ratti, M.~A.~Thaler and W.~Weise,
  Phys.\ Rev.\ D {\bf 73}, 014019 (2006).  

\bibitem{Mocsy:2003tr}
\'A.~M\'ocsy, F.~Sannino and K.~Tuominen,
Phys.\ Rev.\ Lett.\  {\bf 91}, 092004 (2003);
JHEP {\bf 0403}, 044 (2004).

\bibitem{Bernard:2003dk}
C.~Bernard {\it et al.}  [MILC Collaboration],
Nucl.\ Phys.\ Proc.\ Suppl.\  {\bf 129}, 626 (2004);
C.~Bernard {\it et al.}  [MILC Collaboration],
Phys.\ Rev.\ D {\bf 71}, 034504 (2005).

\bibitem{Petreczky:2004pz}
  P.~Petreczky and K.~Petrov,
  Phys.\ Rev.\ D {\bf 70}, 054503 (2004).

\bibitem{Schmidt:2006ku}
  C.~Schmidt,
  hep-lat/0606020.

\bibitem{Schmidt:2006vz}
  C.~Schmidt and T.~Umeda  [RBC-Bielefeld Collaboration],
  hep-lat/0609032.

\bibitem{Petrov:2006pf}
  K.~Petrov,
  hep-lat/0610041.
      
\bibitem{next}
  E. S. Fraga and \'A. M\'ocsy, in preparation.

\bibitem{Lenaghan:2000ey}
J.~T.~Lenaghan, D.~H.~Rischke and J.~Schaffner-Bielich,
Phys.\ Rev.\ D {\bf 62}, 085008 (2000).

\bibitem{Roder:2003uz}
D.~Roder, J.~Ruppert and D.~H.~Rischke,
Phys.\ Rev.\ D {\bf 68}, 016003 (2003).

\bibitem{Pisarski:2000eq}
R.~D.~Pisarski,
Phys.\ Rev.\ D {\bf 62}, 111501 (2000).

\bibitem{Dumitru:2000in}
A.~Dumitru and R.~D.~Pisarski,
Phys.\ Lett.\ B {\bf 504}, 282 (2001).

\bibitem{Kaczmarek:1999mm}
O.~Kaczmarek, F.~Karsch, E.~Laermann and M.~Lutgemeier,
Phys.\ Rev.\ D {\bf 62}, 034021 (2000).
    
\bibitem{matrix}
A.~Dumitru, Y.~Hatta, J.~Lenaghan, K.~Orginos and R.~D.~Pisarski,
Phys.\ Rev.\ D {\bf 70}, 034511 (2004); 
A.~Dumitru, J.~Lenaghan and R.~D.~Pisarski,
Phys.\ Rev.\ D {\bf 71}, 074004 (2005); 
A.~Dumitru, R.~D.~Pisarski and D.~Zschiesche,
Phys.\ Rev.\ D {\bf 72}, 065008 (2005).

\bibitem{Pisarski:2006hz}
  R.~D.~Pisarski,
  hep-ph/0608242.

\bibitem{dissipation}
  E.~S.~Fraga and G.~Krein,
  Phys.\ Lett.\ B {\bf 614}, 181 (2005); 
  E.~S.~Fraga, G.~Krein and R.~O.~Ramos,
  AIP Conf.\ Proc.\  {\bf 814}, 621 (2006); 
  A.~J.~Mizher, E.~S.~Fraga and G.~Krein,
  hep-ph/0604123; 
  L.~F.~Palhares, E.~S.~Fraga, T.~Kodama and G.~Krein,
  hep-ph/0604155; 
  E.~S.~Fraga, T.~Kodama, G.~Krein, A.~J.~Mizher and L.~F.~Palhares,
  hep-ph/0608132.
  
\bibitem{combined}
  E. S. Fraga, A. J. Mizher and \'A. M\'ocsy, work in progress.

 
\end{thebibliography}
\end{document}